\documentclass[namedreferences]{solarphysics}
\usepackage[hyperref,optionalrh,solaromanenum]{spr-sola-addons} 
\usepackage{graphicx}                    
\usepackage{color}                       

\begin{document}

\begin{article}

\begin{opening}

\title{Coronal magnetic field value radial distributions obtained by using the information on fast halo coronal mass ejections}

%
\author[addressref={iszf},corref,email={vfain@iszf.irk.ru}]{\inits{}\fnm{V.G.}\lnm{Fainshtein}}
\author[addressref={iszf},corref,email={egorov@iszf.irk.ru}]{\inits{}\fnm{Ya.I.}\lnm{Egorov}}

%
\runningauthor{Fainshtein et al.}

\address[id=iszf]{Institute of Solar-Terrestrial Physics SB RAS, PO Box 291, Irkutsk, Russia}

\begin{abstract}
Based on the method of finding coronal magnetic field value radial profiles $B(R)$ described in \citep{Gopalswamy2011}, and applied for the directions close to the sky plane, we determined magnetic field value radial distributions along the directions close to the Sun-Earth axis. For this purpose, by using the method in \citep{Xue2005}, from the SOHO/LASCO data, we found 3D characteristics for fast halo coronal mass ejections (HCMEs) and for the HCME-related shocks. Through these data, we managed to obtain the $B(R)$ distributions as far as $\approx43$ solar radii from the Sun center, which is approximately by a factor of 2 farther, than those in \citep{Gopalswamy2011}. We drew a conclusion that, to improve the accuracy of the Gopalswamy-Yashiro method to find the coronal magnetic field, one should develop a technique to detect the CME sites that move in the slow and in the fast solar wind. We propose a technique to select the CMEs, whose central (paraxial) part moves, indeed, in the slow wind.
\end{abstract}

%

\end{opening}

%
\section{Introduction}
     \label{S-Introduction} 

The coronal plasma is immersed in the non-uniform and anisotropic magnetic field. The corona structure, as well as explosive, eruptive processes occurring in it, are closely related to the magnetic field characteristics on different spatial scales. Thereby, developing reliable and precise methods to determine the coronal magnetic field is one of the key problems that solar scientists face.

There have been no regular measurements of the coronal magnetic field; only individual ones based on the Zeeman effect (in the infra-red band; \citealp{Lin2000}), on the Hanle effect \citep{Sahal1986}, on the Faraday effect \citep{Patzold1987}, as well as on the measurements of the solar radio emission \citep{Bogod2016}. The coronal magnetic field may be inferred within different approximations, by measuring the photospheric field: in the potential approximation (see \citep{Rudenko2001} and references therein), in the force-free approximation \citep{Wiegelmann2008,Rudenko2009}.

Individual estimates of the coronal magnetic field were obtained by taking into account the field relation to different characteristics of the coronal plasma. Those characteristics were determined independently, for example, plasma beta, gyrofrequency, Alfv\'en velocity (see the monograph by \citep{Schwenn1990} and references therein). 

In \citep{Gopalswamy2011,Kim2012}, a new method to find the coronal magnetic field values was proposed and tested. The method is based on the relation between the normalized stanoff distance (the distance between the CME part most remote from the Sun, or the CME "nose," to the CME-related shock) and the Alfv\'en Mach number, $M$ \citep{Russell2002}. As consequence this is the relation between the normalized standoff distance and the Alfv\'en velocity. In \citep{Kim2012}, the relation between the Mach number and the shock front density jump value \citep{Landau_EN} was also used to find MA. The method by \citep{Gopalswamy2011,Kim2012} was applied to find the magnetic field radial distributions, $B(R)$, in the sky plane up to approximately $20R_s$ (Rs being solar radius) from COR2 coronagraph data. Those coronagraphs are a part of the Sun Earth Connection Coronal and Heliospheric Investigation (SECCHI) tool suite aboard the Solar Terrestrial Relations Observatory (STEREO; \citealp{stereo}) mission, and from the С2 and С3 Large Angle and Spectrometric Coronagraph (LASCO; \citealp{lasco}) telescopes aboard the Solar and Heliospheric Observatory (SOHO) mission. To find $B(R)$ in the latter case, analyzed were "limb" CMEs, i.e., the mass ejections, whose sources were relatively close to the solar limb. This method in [Poomvises et al., 2012] was applied up to $\approx$120 Rs by using the SOHO/LASCO data and the data from the Heliospheric Imager 1 telescopes within the SECCHI suite. In \citep{Schmidt2016},  for the 2013 November 29 CME, compared were the magnetic field radial distributions obtained by using the method by \citep{Gopalswamy2011} and the magneto-hydrodynamic 3D calculations for the motion of a shock-related model CME. The authors drew a conclusion about a good agreement between the magnetic field radial distributions obtained by two methods up to $(1.8 - 10) ~R_s$.

In this study, we applied the \citealp{Gopalswamy2011} method to find the magnetic induction value in the outer corona to fast halo coronal mass ejections (HCMEs), the sources of most of which are located near the solar disk center. Such mass ejections move at a small angle to the Sun-Earth axis \citep{Fainshtein2006}. Therefore, to find the kinematic characteristics of the CME body and of the shock necessary for determining the magnetic field distribution along the Sun-Earth axis, one should perform special 3D calculations of the CME characteristics. To find the position and the velocity of the HCME body and the related shock boundary in 3D, we used the «ice-cream cone model» addressed in \citep{Xue2005}. This enabled to increase (almost by a factor of 2) the distance, within which the magnetic field value was determined, as compared with that, within which the magnetic field value was determined, when using the limb CMEs from LASCO. The comparison between the radial distributions of the magnetic field obtained by using HCME and the field distributions from limb CMEs in \citep{Gopalswamy2011,Kim2012} may be used as a test for the determining accuracy of the «ice-cream cone model,» when finding the kinematic characteristics of the HCME body and the HCME-related shock in 3D space. The calculated radial field distributions along the Sun-Earth path can be used for MHD calculations of the solar wind and of the model CMEs propagating in it, including the prognostic purposes. And, finally, another aspect (important in our opinion) should be given more light. In \citep{Gopalswamy2011}, the CME central paraxial part is assumed to move in the slow solar wind. In fact, generally, it is difficult to determine, which CME part moves in the slow wind, and which does the same in the fast. The CME may appear to move that, at a sufficient distance from the Sun, a mass ejection may entirely move in the fast wind. We note that, in order to improve the accuracy of the magnetic field radial component determination through the method by \citep{Gopalswamy2011}, one should develop techniques to determine the CME sites moving in the slow and in the fast solar wind. As a first step to solve this problem, in this study, we propose a technique to select the CMEs, whose central parts move in the slow wind.

\section{Data and research techniques}
     \label{S-Data} 

For the analysis, we selected the following fast HCMEs: 2003 Nov 18 (8:50:05 UT), 1660 km/s, N00E18, M3.9; 2004 Apr 06 (13:31:43 UT), 1368 km/s, S18E15, M2.4; 2004 Nov 03 (16:54:05 UT), 1759 km/s, N09W17, X2.0; 2004 Nov 07 (16:54:05 UT), 1759 km/s, N09W17, X2.0; 2005 Jan 15 (23:06:50 UT), 2861 km/s, N15W05, X2.6; 2005 Jan 17 (9:30:05 UT), 2094 km/s, N13W19, X2.2; 2005 Jul 30 (6:50:28 UT), 1968 km/s, N12E60, X1.3; 2005 Sep 05 (19:48:05 UT), 2257 km/s, source beyond the limb; 2005 Sep 13 (20:00:05 UT), 1866 km/s, S09E10, X1.5. For each event, we provide the time (in brackets) of the first HCME recording within the LASCO C2 field-of-view, then, we show the linear projection velocity of the mass ejection, the coordinates and the X-ray class of the HCME-related flare.

To find the HCME 3D parameters, we used the method proposed in \citep{Xue2005}, where the so-called «ice-cream cone» model was used as an ejection model. In this model, a coronal mass ejection is represented like a cone with the top at the Sun center. The cone leans against the part of the spherical surface with the radius equal to the cone generator length. The HCME motion direction is determined by the position of the model cone axis. This direction is described by two angles: $\theta_0$ and $\phi_0$. The $\theta_0$ angle (colatitude) is counted ($\theta_0$ = $[0^\circ; 180^\circ]$) from the positive direction of the axis transiting through the Sun center to the north and perpendicular to the ecliptic plane (in this work we neglect difference between the plane of the solar equator and the plane of an ecliptic). The longitudinal angle $\phi_0$ is counted off in the ecliptic plane from the central meridian counterclockwise ($\phi_0$ = $[0^\circ; 360^\circ]$). Except the motion direction, this method enables to calculate its motion velocity, vp on the model CME axis, as well as the mass ejection angle size α. At first sight, this is a very simple CME model, and, to solve the set problem, it was worth using more realistic models for coronal mass ejections. In \citep{Michalek2006}, used was the CME cone model considered more realistic, with an elliptical shape of the model cone basis. \citealp{Thernisien2006} used an even more realistic CME model like a magnetoplasma rope. The results in \citep{Xue2005} enable to regard possible using the cone model to calculate CME 3-D parameters and the CME-related shock \citep{Kim2011ab}. In that study, the authors showed  that the CME radial velocities calculated by using the CME models proposed by \citep{Michalek2006,Thernisien2006} are very close to the velocities calculated through the method from \citealp{Xue2005}, with the correlation coefficient being more, than 0.95.

In \citep{Xue2005}, when calculating the fast CME 3D parameters, the authors did not take into account that the moving higher brightness regions observed within the LASCO C2 and C3 fov (identified as CMEs), in fact, involve the CME body and the CME-related shock. There is shock-compressed plasma between the CME body and the shock. We applied the method by \citep{Xue2005} to calculate 3-D parameters separately for the CME body and for the shock. To find the magnetic induction values, we modified this method to calculate not only the velocities of the CME body and the shocks along their axes, but also the positions of these structures along their axes relative to the Sun center.

Our implementing the method proposed by \citep{Gopalswamy2011} is in the following.

1) For each addressed HCME at different instants of its motion within the LASCO C3 fov, we calculate the distance ΔR between the shock and the CME body along the axis of the model mass ejection by using the method from \citep{Xue2005}, as well as the CME-body boundary curvature radius Rc.  Note that, unlike \citep{Xue2005}, where used were the coordinates of a discrete set of shock points to calculate the model CME 3D parameters, we used delineation both the shock and the CME body by ellipse segments (Fig. 1).

2) From the $\Delta R/R_c = 0.81 [(\gamma - 1) M^2 + 2]/[(\gamma + 1) (M^2 - 1)]$ relation \citep{Russell2002}, we find the Alfv\'en Mach number, $M$ ($\gamma$ was considered 4/3).

3) From the $M = (V_{Sh}-V_{SW})/VA$ formula, we find the Alfv\'en velocity, $V_A$. Here, $V_{Sh}$ is the shock velocity, $V_{SW}$ is the velocity of the solar wind, through which the shock propagates. Like in \citep{Gopalswamy2011}, we assume that the CME bulk moves in the region of the slow solar wind, whose velocity we find from the $V^2_{SW} (R) = 1.75 \times 10^5 (1 - exp (- (R - 4.5)/15.2))$ relation \citep{Sheeley1997}.

4) From the $V_A = 2.18 \times 10^6n^{-1/2}B$ formula, we find the value for the magnetic induction $B$ (in G). In this formula, $n$ (in cm$^{-3}$) is the coronal plasma particle concentration that is assumed to be equal to the electron concentration, and is found from the $n(R) = 3.3 \times 10^5R^{-2} + 4.1 \times 10^6R^{-4} + 8.0 \times 10^7R^{-6}$ relation \citep{Leblanc1998}. Here, $R$ is the distance in the sky plane from the solar disk center to the observation point.

\section{Results}
     \label{S-Results} 

Figure 1 provides an example of an HCME within the LASCO C3 fov with the CME body and shock boundaries delineated by ellipse segments. Note that the delineating of the indicated structures was not always performed with their complete latitude cover within $360^\circ$. We delineated only those sites of the structures that we could robustly identify as an HCME body or a shock. Our analysis showed that the difference in the 3D parameters calculated for the same events between incomplete and complete ($360^\circ$) delineates is relatively insignificant.

The diffuse region boundary was considered a shock, because the velocity of this boundary relative to the ambient slow solar wind exceeded the Alfv\'en velocity. At separate segments of this boundary, on the brightness scans along the directions perpendicular to the boundary, one can detect brightness jumps with the $(1-2)~\delta R$ spatial size, where $\delta R \approx 0.125R_s$ is the LASCO C3 spatial resolution, $R_s$ is the Sun radius. Such shocks are referred to as collisionless \citep{Artsimovich1979}, because their front width is manifold less, than the mean length of proton and electron collisions with the coronal plasma protons, $L_c (L_c \approx (1-3) R_s$ within the LASCO C3 fov). The real front width of a collisionless shock, according to [Artsimovich and Sagdeev, 1979], and taking into account the characteristics of the coronal plasma and of the magnetic field, is manifold less, than $\delta R$.

\begin{figure}[!ht] 
\centerline{\includegraphics[trim=0.0cm 0cm 0.5cm 0cm, width=1.0\textwidth]{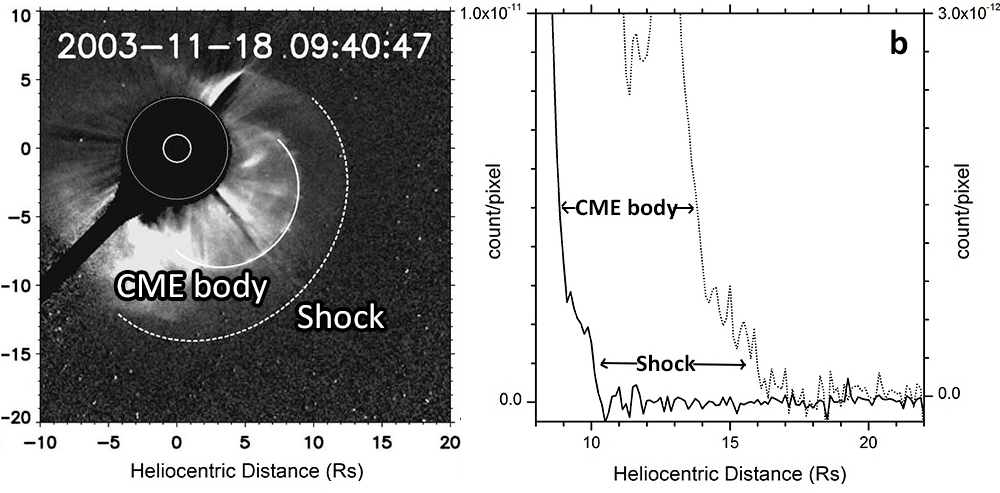}}
\caption{(a) HCME recorded 2005 Jan 15 (23:41:27 UT). The white ellipses show the boundaries of the mass ejection body (bright inner domain) and of the shock (outer diffuse domain boundary). (b) The brightness distributions along 30NW for two instants. Well-defined are the brightness jumps with a spatial size $(1-2)~\delta R$ spatial size, where $\delta R \approx 0.125R_s$ is the LASCO C3 spatial resolution, $R_s$ is the Sun radius. The shock is assumed to be collisionless. }
\end{figure} 

Figure 2 presents the magnetic field calculation results obtained by using the HCME 3D parameter calculation. From this figure, it follows that the obtained $B(R)$ dependence agrees quite well with the expected values of the slow solar wind magnetic field.

\begin{figure}[!ht] 
\centerline{\includegraphics[trim=0.0cm 0cm 0.5cm 0cm, width=1.0\textwidth]{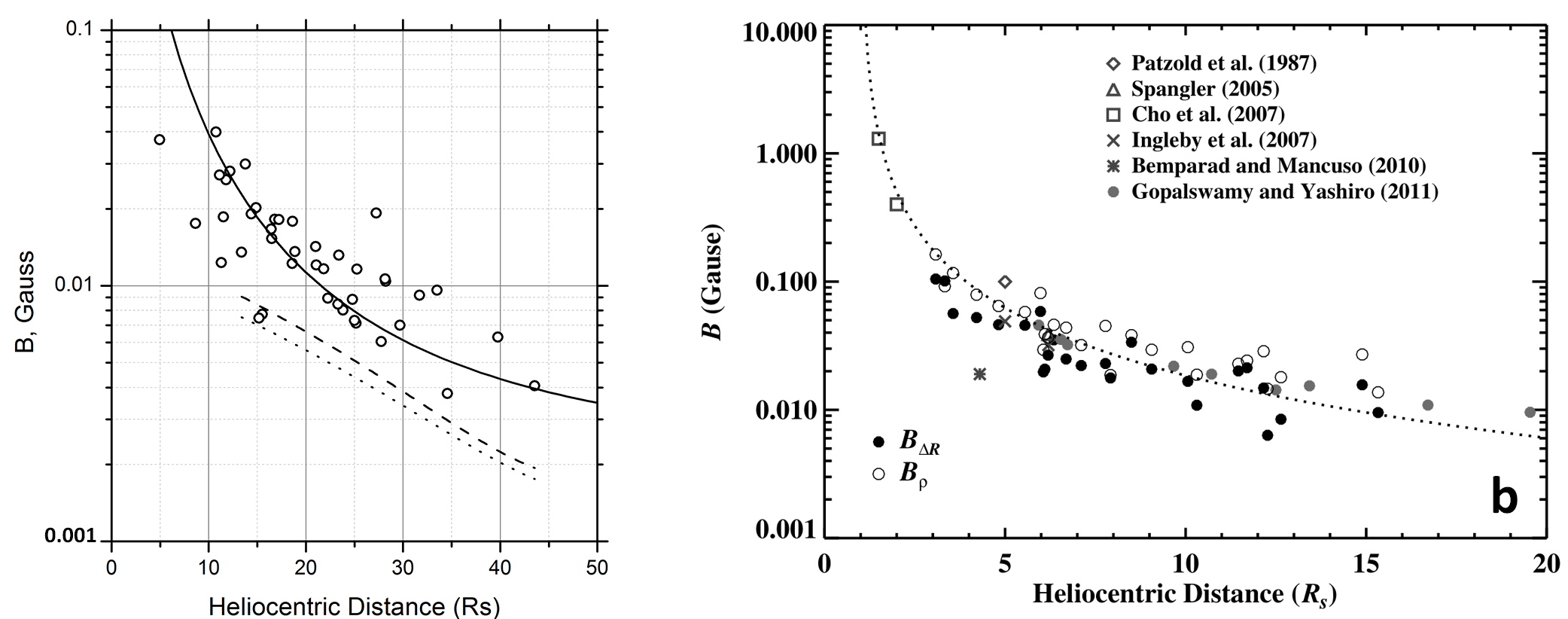}}
\caption{Left panel: circles present the $B(R)$ values obtained by the HCME 3D parameter calculations at different instants for each of the addressed event. The thick line crossing the circles is the $B_r(R) = 8\times10^{-5}(215.5/(R/R_s))^2 +0.002$, G is the mean value of the magnetic field radial component in a horizontal site of the slow solar wind \citep{Fainshtein1991} depending on the distance. The 0.002 summand is a correcting quantity to improve the correspondence between the $B_r(R)$ and the $B(R)$ measured values dispersion. Two lines below: field calculations under the assumption that the CME paraxial region moves in the fast solar wind (SW), whose velocity $V_{SW}=600~ km/s$ (upper curve) and 800 km/s (lower curve) irrespective of the distance, and with the proton concentration dependence on the distance, characteristic of the fast SW: $N_p = 3\times(215.5/(R/R_s))^2$. Here, we present only the regression lines (see the discussion of these curves see in the Conclusion). Right panel: Fig. 8 from \citep{Kim2012} (© AAS. Reproduced with permission). In that figure, solid circles are the $B(R)$ values obtained from the relation associating the Mach number with $\Delta R/R_c$ ($B\Delta R$), open circles are the $B(R)$ values obtained from the relation associating the Mach number with the shock front jump density ($B \rho$). Other symbols show the values for the magnetic induction obtained by different authors through other methods. We do not give the references to the papers, to which these symbols relate.}
\end{figure} 

It is not difficult to realize that, within (10-20) $R_s$, the difference between our results and those of \citep{Kim2012} is not great: on average, our $B(R)$ values are by factor of $\approx1.5$ more than the field values obtained by \citealp{Kim2012}.

Figure 3 on the panels in the upper row illustrates the distance dependence of the Alfv\'en velocity $V_A$. The left panel shows the dispersion of the $V_A$ values obtained from the HCME data. The right panel presents the $V_A$ distance dependences that we calculated for the slow solar wind (lower strip/band) and for the fast wind (upper strip/band). In this case, we used the magnetic field values and the proton concentration in the Earth orbit and certain dependences of their variation with distance. The same figure shows (in the lower row) the $V_A$ dependence on the shock position from \citep{Kim2012}.

\begin{figure}[!ht] 
\centerline{\includegraphics[trim=0.0cm 0cm 0.5cm 0cm, width=1.0\textwidth]{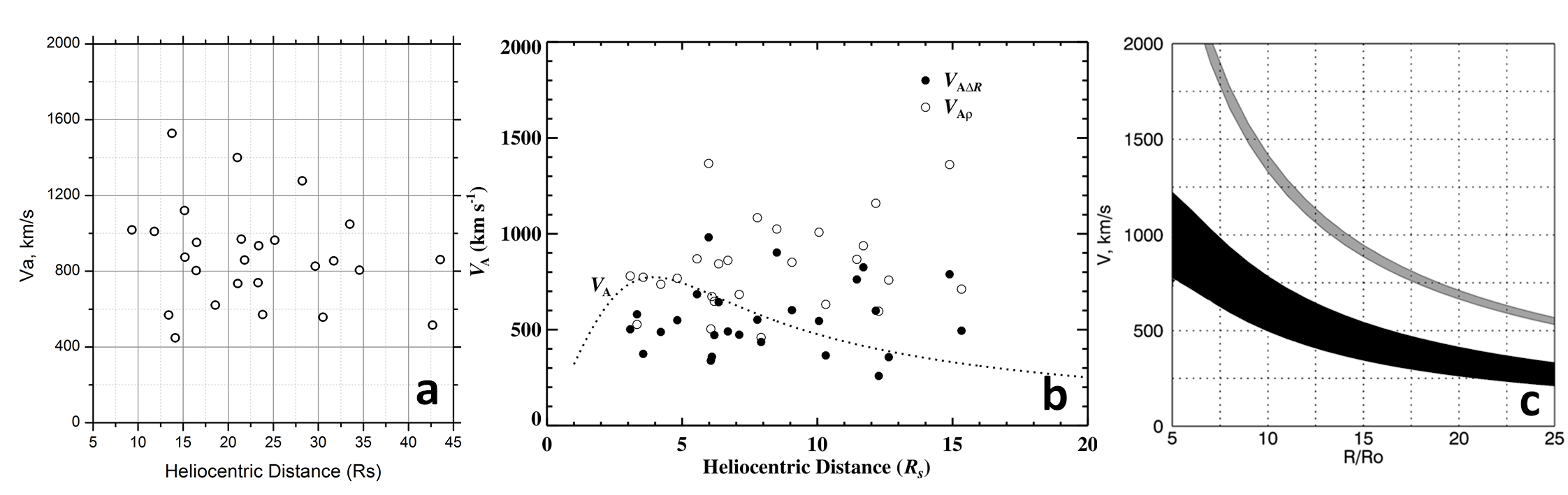}}
\caption{Left panel in the upper row: distance-dependant dispersion of dots ($V_A$ values). The $V_A$ values were obtained in 3D from the HCME data. The right panel presents the $V_A(R)$ dependence obtained for the slow solar wind (lower strip/band) and for the fast wind (upper strip/band) by using the magnetic field values and the proton concentration in the Earth orbit and certain dependences of their variation with distance. Dots show the $V_A(R)$ dependence for quiet regions of the corona from \citep{Mann1999}. The panel in the lower row is Fig. 7 from \citep{Kim2012} (© AAS. Reproduced with permission). The dotted line provides the $V_A(R)$ dependence from \citep{Mann1999}. In that figure, solid circles are the $B(R)$ values obtained from the relation associating the Mach number with $\Delta R/R_c$, while open circles are the $B(R)$ values obtained from the relation of the density at the front shock.}
\end{figure} 

From Figure 3, one can see that, on average, within (5-10) $R_s$, the Alfv\'en velocities obtained from the HCME data are approximately by 100-200 km/s higher than those in Figure 7 from \citep{Kim2012}. At the same time, the plot built from the HCME data demonstrates more clearly the Alfv\'en velocity decrement with distance. Note also that the $V_A$ value spread for each shock position within up to 15 $R_s$, obtained both in our calculations and in \citep{Kim2012}, is essentially more as compared with the Alfv\'en velocity calculations from the magnetic field and the plasma density data in the Earth orbit. We assume that this evidences a lower accuracy when calculating the Alfv\'en velocity by using the methods proposed in \citep{Gopalswamy2011,Kim2012}.

Figure 4 (left panel in the upper row) shows the dispersions of the dots (the Alfv\'en Mach number values for different shock positions) that we obtained. The right panel in the upper row shows a similar result from (\citep{Kim2012}, their Fig. 6). One can see that, within (8-15) $R_s$, the spread of the Mach number values and the mean M value for the two types of calculations are close.

In the lower row on that figure, we show (left panel) the dispersion of our $\Delta R/R_c$ values for different shock positions, and (right panel) the dispersion of the same parameter from \citep{Kim2012}, their Fig. 2). Within (8-20) $R_s$, the $\Delta R/R_c$ minimal values in two cases are close. The maximal value for this parameter is $\approx$ 0.75 and $\approx$ 0.6, the mean value being $\approx$ 0.3 and $\approx$ 0.4. Note that the $\Delta R/R_c$ parameter decreases, on average, as the shock moves away.

\begin{figure}[!ht] 
\centerline{\includegraphics[trim=0.0cm 0cm 0.5cm 0cm, width=1.0\textwidth]{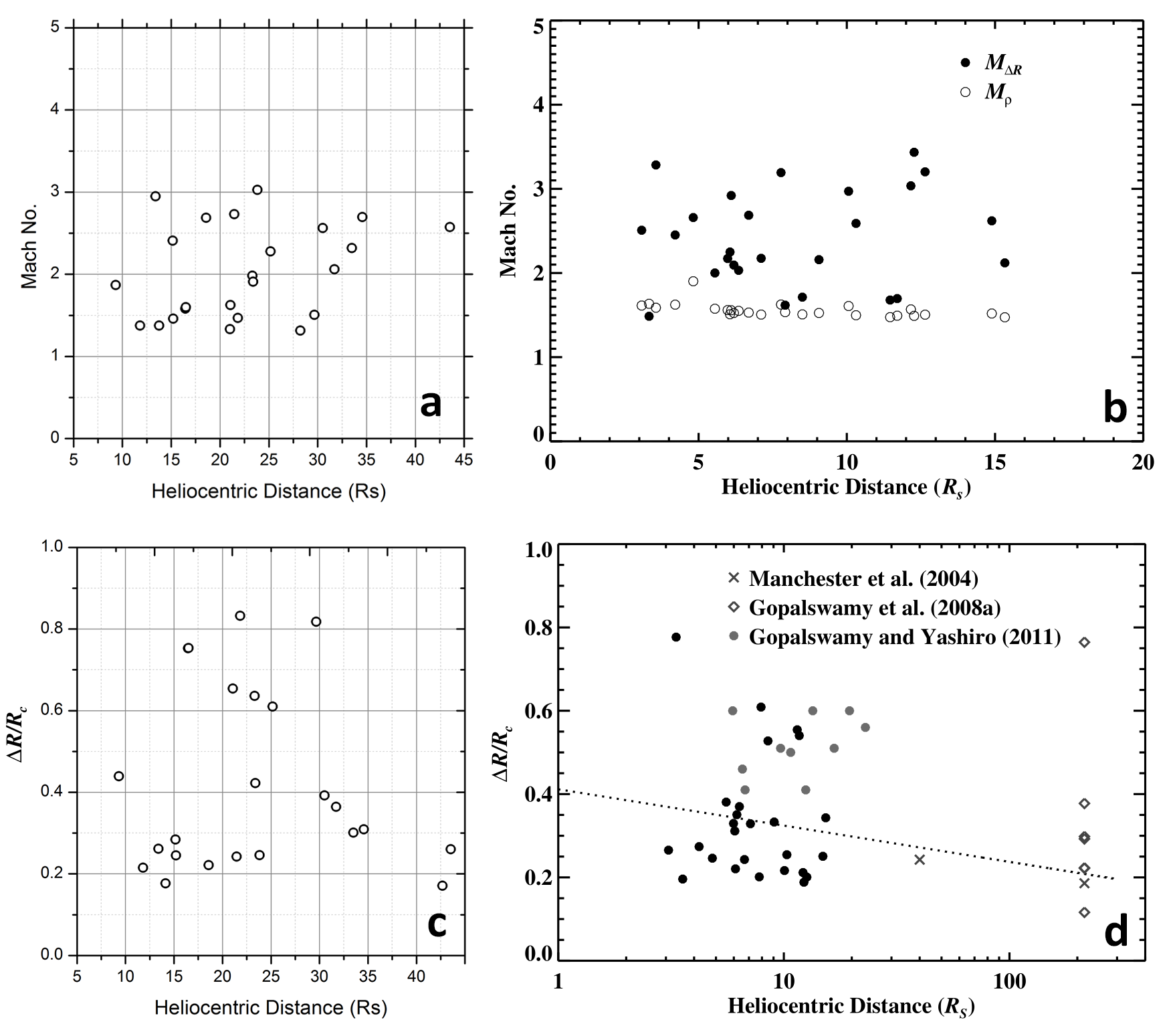}}
\caption{Upper row of panels: left - dispersion of the Mach number values that we obtained for different shock positions; right - Figure 6 from \citep{Kim2012} ((© AAS. Reproduced with permission)). The lower row of panels: left - dispersion of our $\Delta R/R_c$ values; right - Figure 2 from \citep{Kim2012} (© AAS. Reproduced with permission).}
\end{figure} 

\newpage

\section{Discussion and conclusions}
     \label{S-Conclusion} 

In this study, we obtained spatial distributions of the magnetic field B(R) within the (5-43) Rs shock positions approximately along the Sun-Earth axis, by using the method proposed in \citep{Gopalswamy2011}. We found the CME body curvature radius in the surroundings of its axis and the distance between the CME body boundary and the CME-related shock in 3D (both necessary for this purpose) through the CME «ice-cream cone model» \citep{Xue2005}. The obtained distribution B(R) is close to the radial distribution of the magnetic induction obtained in \citep{Kim2012} in the sky plane for limb CMEs. On average, this distribution is also close to the dependence of the field radial component varying with distance by the law $B_r = B_{re}(215.5/(R/R_s))^2$, where $B_{re}$ is the mean value $B_r$ in a horizontal site of the slow solar wind in the Earth orbit (see below about horizontal sites of the slow wind). We also compared our values for the Alfv\'en velocity, Alfv\'en Mach number $M$, and the $\Delta R/R_c$ parameter (where $\Delta R$ is the distance between the shock and the CME body along the model mass ejection axis, $R_c$ is the CME body boundary curvature radius) with those in \citep{Kim2012}. Here, one may also note the precision of both mean values ($\approx$0.32 for $\Delta R/R_c$; $\approx$2 for $M$), and $M$ and $\Delta R/R_c$ spreads relative to the mean values at different shock positions ($R$). Based on the obtained results, we can draw a conclusion that the "ice-cream cone model" used to calculate the CME 3D parameters suits well for calculating the magnetic field along the CME axis in 3D, including the position along the Sun-Earth axis.
In \citep{Gopalswamy2011} and in our calculations of the magnetic field value, the paraxial sites of all the CMEs, whose properties were used to find the magnetic field value, were implicitly supposed to move in the slow solar wind region. It is this reason, why we used the distance-dependant slow wind velocity from \citep{Sheeley1997} as the velocity of the solar wind where a shock propagates. In fact, for many CMEs, it is sufficiently difficult to specify, which CME parts move in the slow wind. Moreover, in certain cases, the entire CME may move (within the LASCO C3 fov) in the fast solar wind, and, at high velocities (for example, higher than 1500 km/s), there may exist a shock ahead of it. Assuming that, in all the addressed events, the CME moves in the fast solar wind, we found the magnetic field radial distribution by using the method from \citep{Gopalswamy2011}. Figure 2 shows the regression lines for the magnetic induction dependence on the shock position. Apparently, in this case, the magnetic field is essentially weaker, than that in the slow wind.

Figure 5 shows the WSO-calculated NL with the segments located at small angles to the solar equator plane (i.e., almost parallel to that plane), and the CME moving along a bright streamer corresponding to one of such segments. This implies that its central part moves in the slow solar wind region.

\begin{figure}[!ht] 
\centerline{\includegraphics[trim=0.0cm 0cm 0.5cm 0cm, width=1.0\textwidth]{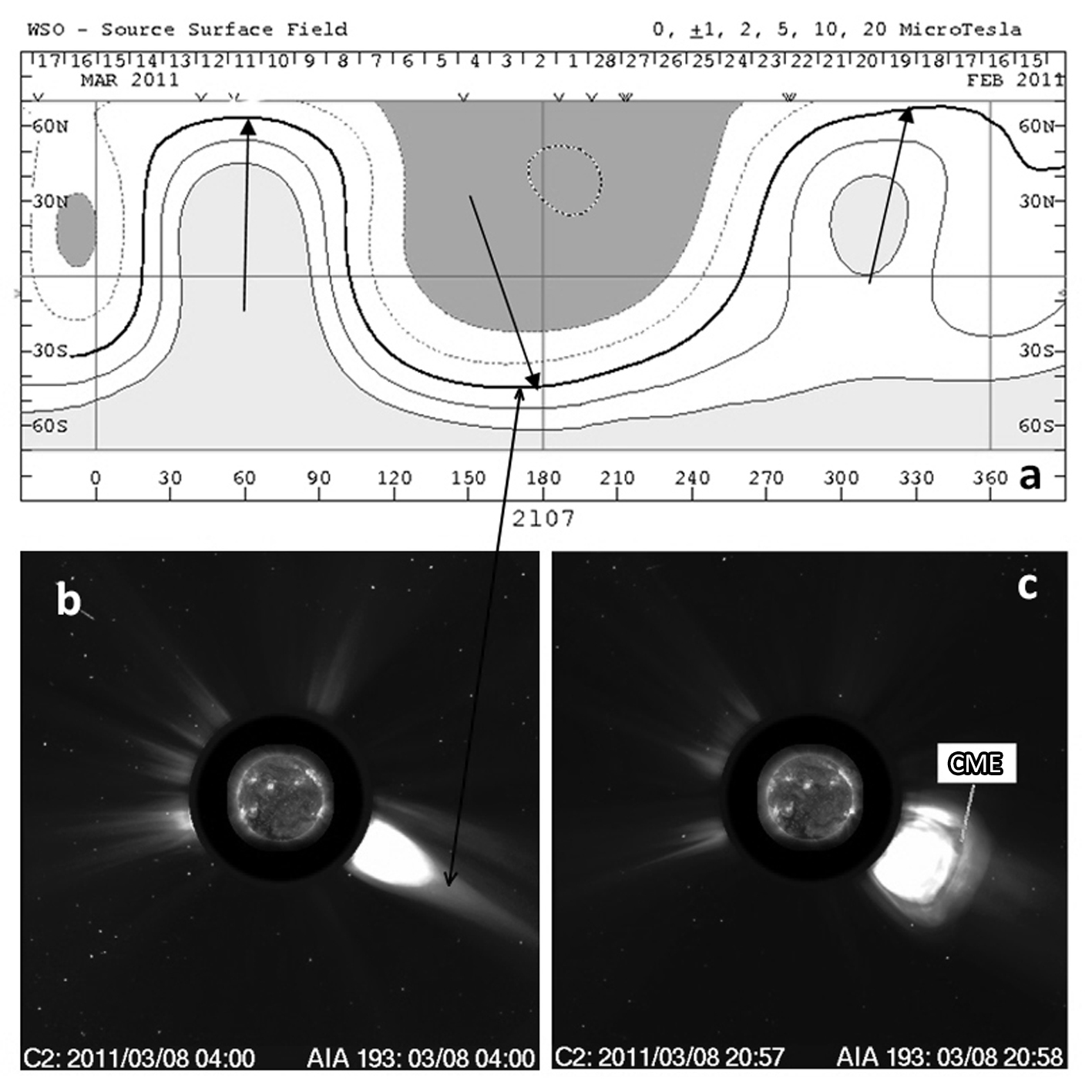}}
\caption{Top panel shows the magnetic field distribution on the source surface ($R_{ss}=2.5R_s$) from the field calculations in potential approximation at WSO (\url{http://wso.stanford.edu/synoptic/WSO-S.2107.gif}). The arrows mark the NL segments located at small angles to the solar equator plane. The CME effect on the streamer led to its brightness decrease ahead of the CME. The left panel in the lower row is the white corona (the region beyond the artificial, i.e., the black ring). The green color domain is the Sun image in the 19.3 nm channel from SDO. The red arrow shows the coronal streamer before the CME emergence and the NL segment corresponding to this streamer. The right panel shows the CME, whose middle part moves through the coronal streamer shown on the left panel.}
\end{figure}

%

%
 \begin{acks}
The authors are grateful to the LASCO team for a possibility to freely use the coronagraph data. The study was done within the ISTP SB RAS 2016-2018 R\&D Plan II.16.1.6 entitled «Geoeffective Processes in the Sun Chromosphere and Corona» (basic project), and, partially, with a support from the Russian Foundation for Basic Research Grants No. 15-02-01077-а and No. 16-32-00315.
 \end{acks}

%
%
\newpage

 \bibliographystyle{spr-mp-sola}
 \bibliography{biblio2}

\end{article} 
\end{document}